\documentclass{article}
\usepackage{spconf,amsmath,graphicx,hyperref}
\usepackage{cite}
\usepackage{amssymb,amsfonts}
\usepackage{algorithmic}
\usepackage{graphicx}
\usepackage{textcomp}
\usepackage{xcolor}
\usepackage{tabularx}
\usepackage{multirow}
\usepackage{booktabs}
\usepackage{makecell}

\title{Generalizability of Predictive and Generative \\ Speech Enhancement Models to Pathological Speakers}
\name{Mingchi Hou$^{1 ,2}$, Ante Juki{\'c}$^{3}$, Ina Kodrasi$^{1}$\thanks{This work was supported by the Swiss National Science Foundation project 200021\_215187 on ``Pathological Speech Enhancement".}}
\address{$^{1}$Idiap Research Institute, Switzerland \\$^2$École Polytechnique Fédérale de Lausanne, Switzerland \\ $^{3}$NVIDIA, USA}

\def\ninept{\def\baselinestretch{.941}\let\normalsize\small\normalsize}

\begin{document}
\ninept

\maketitle

\begin{abstract}
State-of-the-art speech enhancement (SE) models achieve strong performance on neurotypical speech, but their effectiveness is substantially reduced for pathological speech. In this paper, we investigate strategies to address this gap for both predictive and generative SE models, including (\textit{i}) training models from scratch using pathological data, (\textit{ii}) fine-tuning models pre-trained on neurotypical speech with additional data from pathological speakers, and (\textit{iii}) speaker-specific personalization using only data from the individual pathological test speaker. Our results show that, despite the limited size of pathological speech datasets, SE models can be successfully trained or fine-tuned on such data. Fine-tuning models with data from several pathological speakers yields the largest performance improvements, while speaker-specific personalization is less effective, likely due to the small amount of data available per speaker. These findings highlight the challenges and potential strategies for improving SE performance for pathological speakers.
\end{abstract}


\section{Introduction}
Speech communication is essential for human interactions, facilitating information exchanges across diverse environments. However, in real-world settings, background noise often degrades speech quality and intelligibility. Speech enhancement (SE) techniques aim to address this challenge by estimating clean speech from noisy recordings.
The widespread adoption of voice-based technologies, such as mobile communication, hearing aids, automatic speech recognition, and teleconferencing systems, alongside the need to alleviate cognitive burden and ease listener’s fatigue, has increased the demand for robust and generalizable SE solutions~\cite{oshaughnessy_speech_2024}. This need drives efforts like the URGENT challenge~\cite{zhang_urgent_2024} for universal SE robustness across languages, noises, and domains.

Traditional SE approaches, such as spectral subtraction~\cite{boll_suppression_1979} or Wiener filtering~\cite{lim_enhancement_1979}, rely on assumed statistical properties of the clean speech and noise signals.
While effective in stationary noise environments, these methods often struggle with non-stationary interferences. 
The advent of deep learning has revolutionized the field, with approaches employing data-driven solutions to model complex signal patterns from training data~\cite{wang_overview_2018,pascual_segan_2017}.
Data-driven solutions have traditionally relied on predictive models that aim to learn a single deterministic mapping from noisy to clean speech, for example by estimating spectral masks~\cite{CNNGRU_mask, NTT2020} or spectral coefficients~\cite{Sun_2017}.
Such approaches, however, encounter challenges to generalize to unseen noise environments~\cite{generalization_gap2023}. 
More recently, the emergence of generative models has become an increasingly popular area of research in various speech related tasks, including SE~\cite{richter_speech_2023, lu_diffusion_ddpm_2021, lu_cdiffuse_2022}. 
State-of-the-art (SOTA) generative SE approaches are typically based on diffusion models, operating by progressively adding noise to a clean speech signal in a forward diffusion process~\cite{lu_diffusion_ddpm_2021}. The reverse process, which can be guided by a neural network, learns to recover clean signals from noisy inputs.
Recently, a Schr\"{o}dinger bridge-based generative model was proposed for SE~\cite{jukic2024sb}.
Differently from typical forward diffusion, this model results in exact interpolation between the clean and noisy speech spectral coefficients.

Despite beneficial enhancement performance, the vast majority of SE research is using recordings from neurotypical speakers exhibiting no speech disorders from datasets such as Wall Street Journal~\cite{richter2023sgmse, jukic2024sb, lemercier_storm_2023} or VoiceBank\cite{DVAE, lemercier_storm_2023}. 
Pathological speech, produced by individuals with hearing impairments, head and neck cancers, or neurological conditions such as Parkinson’s disease (PD), has received far less attention.
Such speech often exhibits atypical acoustic characteristics~\cite{Darley_dysarthria}, leading to reduced intelligibility and increased vulnerability to noise.
Research has further shown that pathological speech differs markedly in its statistical distribution compared to neurotypical speech~\cite{PD_ITG,kodrasi_spectro-temporal_2020}, which implies that SE models trained exclusively on neurotypical data are likely to generalize poorly to pathological conditions.
Although these conditions are widespread, with approximately $360$ million people experiencing hearing impairment~\cite{WHO_2013}, $650$ thousand new cases of head and neck cancers diagnosed annually~\cite{bray_cancer_2018}, and over $1$ billion individuals worldwide affected by neurological disorders~\cite{WHO_2006}, the effectiveness of SOTA SE models for pathological speakers remains largely unexplored.
To the best of our knowledge, SE performance for pathological speech has only been explicitly studied in~\cite{hou_eusipco_2025}, where a hybrid variational autoencoder (VAE)–non-negative matrix factorization (NMF) model was evaluated.
While results showed a marked deterioration in enhancement quality for pathological speakers, this finding is limited in scope since the VAE–NMF model neither reflects the architectures nor the performance levels of current SOTA systems~\cite{richter2024investigating}.
Consequently, the true capability of modern SE models to handle pathological speech remains unknown.
Given the global prevalence of pathological speech, systematically assessing the performance of SOTA SE models and developing tailored solutions for handling pathological speech is therefore a critical research need.

This paper systematically investigates the performance of SOTA predictive and generative SE models on pathological speech and proposes strategies to improve their effectiveness. Specifically, we consider: (\textit{i}) training models from scratch using pathological data, (\textit{ii}) fine-tuning models pre-trained on neurotypical speech with additional pathological data, and (\textit{iii}) speaker-specific personalization where models pre-trained on large neurotypical speech corpora are fine-tuned using only data from the pathological speaker of interest.

\section{Speech Enhancement}
Consider the noisy microphone signal $y(\tau)$ at time $\tau$, i.e.,
\begin{equation}
y(\tau) = x(\tau) + n(\tau),
  \label{equation:eq1}
\end{equation}
with $x(\tau)$ denoting the clean speech signal and $n(\tau)$ denoting the additive noise signal.
In the short-time Fourier transform (STFT) domain, the signal model in~(\ref{equation:eq1}) is given by
\begin{align}
  Y(j,k) = X(j,k) + N(j,k),
  \label{equation:eq3}
\end{align}
where $Y(j,k)$, \(X(j,k)\) and \(N(j,k)\) are the complex-valued STFT coefficients of noisy, clean speech, and noise signals, \(j\) represents the frequency bin index, and \(k\) represents the time frame index. 
The objective of single-channel SE is to estimate the clean speech signal given the noisy microphone recording. 
In the following, several enhancement approaches considered in this paper are briefly reviewed.
We consider predictive models, which learn a single mapping
between noisy and clean speech, as well as generative models, which learn the distribution of clean speech.

\subsection{Predictive models}

\emph{Magnitude spectrogram-based masking (MM):} \enspace Mask-based models estimate the clean speech signal by selectively masking unwanted (i.e., noise-dominated) time-frequency components~\cite{vincent_spectral_2018}.
The enhanced signal is computed as $\hat{X}(j,k) = M(j,k)Y(j,k)$, where $M \in [0,1]$ is the time-frequency mask. 
In this paper, we use a deep neural network (DNN) model trained to estimate the traditional ideal ratio mask, which is defined as the ratio between the spectral magnitudes of the clean and noisy speech~\cite{NTT2020}. 
While various training losses have been investigated for mask-based models for SE~\cite{braun_consolidated_2020}, in this paper we use the widely adopted scale-invariant signal-to-distortion ratio (SI-SDR)~\cite{roux_sdr_2019} between the time-domain predicted signal $\hat{x}(\tau)$, computed using the inverse STFT of $\hat{X}(j,k)$, and the clean reference signal \( x(\tau) \).

\emph{Complex-valued spectrogram-based regression (CR):} \enspace
Instead of enhancing only the noisy speech magnitude and retaining the noisy phase as in the magnitude spectrogram-based masking, approaches that jointly enhance the magnitude and
the phase component of the noisy signal have also become popular. 
For this category of approaches, we use a DNN model trained to estimate the real and imaginary part of the clean STFT coefficients from the noisy STFT coefficients~\cite{song2019generative}.
Since the mean square error (MSE) is a standard and straightforward choice in regression problems, the training loss used for this approach is the MSE between the time-domain predicted signal $\hat{x}(\tau)$ and the clean reference signal $x(\tau)$.

\subsection{Generative models}

{\emph{Score-based diffusion model (SGMSE+):}} \enspace
Diffusion models typically operate by progressively adding noise to a clean speech signal in a forward diffusion process~\cite{song2019generative, song_2021_score, richter2023sgmse, lemercier24spm}. A score-model can be trained to guide the reverse process, and hence, recover clean signals from noisy inputs by removing the noise at each reverse step~\cite{richter2023sgmse, lemercier24spm}. The diffusion process is defined by a forward stochastic differential equation (SDE)~\cite{song2019generative, song_2021_score}. The corresponding reverse SDE can be expressed in terms of the forward SDE parameters with an additional term based on the score function $\nabla_{x} \mathrm{log} \, p_t(x)$ of the marginal distribution $p_t$ at process time $t$. To enable inference using the reverse SDE, a DNN is trained to estimate the score~\cite{richter2023sgmse} and an iterative sampler is used to obtain the clean speech estimate. In this work, an SGMSE+ model incorporating an affine drift term with a predictor-corrector sampler is used.
For additional details, the reader is referred to~\cite{richter2023sgmse}.

\emph{Schr\"{o}dinger Bridge-based model (SB):} \enspace 
Schr\"{o}dinger Bridge is a generative model that aims to find an optimal transport path that minimizes the discrepancy between noisy and clean distributions~\cite{schrodinger1932theorie, chen2023sb, jukic2024sb}. As opposed to typical forward diffusion~\cite{richter2023sgmse}, where clean data is transformed into a sample from a broad distribution centered around the noisy observation, the SB results in exact interpolation between the clean speech and the observed noisy speech coefficients~\cite{jukic2024sb}. Thus, a weighted data prediction loss is used. In this work, a SB model with SDE sampler is used. For additional details, the reader is referred to~\cite{jukic2024sb}.

\section{Pathological SE Strategies}
\label{sec: patho_se}

It is well established that SE models exhibit variable performance due to mismatches in speaker characteristics between training and testing sets~\cite{richter2023sgmse,gonzalez_icassp_2024}. 
This issue is expected to be even more pronounced for pathological speakers, whose acoustic characteristics differ substantially from neurotypical speakers~\cite{PD_ITG,kodrasi_spectro-temporal_2020}. 
To improve the performance of SOTA SE models for pathological speakers, we investigate three strategies.
First, we consider the feasibility of training models from scratch on a (small) dataset that includes both neurotypical and pathological speech. 
Second, we explore fine-tuning models pre-trained on large neurotypical datasets on smaller datasets containing pathological speech. Based on our previous work with the hybrid VAE–NMF model~\cite{hou_eusipco_2025}, fine-tuning with pathological data is expected to improve performance for pathological speakers. However, a performance gap relative to neurotypical speakers may persist due to the high variability in pathological speech characteristics.
Finally, we consider speaker-specific personalization, where the parameters of models pre-trained on large nurotypical speech datasets are adapted to individual test speakers. 
While this approach can provide some gains, its effectiveness is limited in our experiments, likely due to the small amount of data available per speaker. 

\section{Experimental Settings}

\subsection{Neurotypical and pathological datasets}

{\emph{Clean speech datasets:}} \enspace For the results presented in the following section, we use the Spanish CROWD~\cite{guevara-rukoz_crowdsourcing_2020} and PC-GITA~\cite{orozco-arroyave_new_2014_pcgita} clean speech datasets.
CROWD is a large dataset of neurotypical speech used to pretrain the considered SE models. 
It contains $37.8$ hours of recordings from $174$ healthy speakers sampled at $48$ kHz. To reflect the structure of standard SE datasets in terms of duration and number of speakers (cf. e.g.,~\cite{richter2023sgmse, jukic2024sb}), recordings are downsampled to $16$ kHz, and we use $23$ h, $2.2$ h, and $1.5$ h of the data for training, validation, and testing respectively.
PC-GITA contains $2.8$ hours of recordings sampled at $44.1$ kHz from $50$ patients diagnosed with PD and $50$ neurotypical controls, reflecting the small size of pathological speech datasets that are typically available.
Each speaker utters $12$ utterances ($10$ sentences, $1$ read text, $1$ monologue). Recordings are downsampled to $16$ kHz.
Given the small dataset size, we adopt a $10$-fold speaker-independent cross-validation strategy when training or fine-tuning models using the PC-GITA dataset, with $80$\%, $10$\%, and $10$\% of the data used for training, validation, and testing respectively in each fold.

\emph{Noisy mixtures:} \enspace
To generate noisy mixtures, we use the CHiME3 dataset~\cite{barker_chime3_2015}. All noise signals are first downsampled to $16$ kHz. For each clean utterance, we randomly select a noise file recorded on a bus, in a cafe, in a pedestrian area, or at a street junction, and add it at an SNR uniformly sampled between $-6$ dB and $14$ dB for training and validation. For testing, we use fixed SNRs of $-5$ dB, $0$ dB, $5$ dB, $10$ dB, and $15$ dB to provide a consistent evaluation of model performance.

\subsection{Training Configuration}
As in~\cite{lemercier_storm_2023}, signals are transformed to the STFT domain using a window size of $510$ samples and a hop size of $128$ samples. Further, the hyperparameters used to compress the dynamic range of the spectrogram are $\alpha = 0.5$ and $\beta = 0.33$~\cite{jukic2024sb}.

The \emph{MM model} is trained using a $5$ layer Bidirectional Long Short-Term Memory network~\cite{NTT2020}. 
The \emph{CR model} follows the NCSN+ architecture in~\cite{song2019generative}, i.e., a multi-resolution U-Net architecture with skip connections, incorporating $3$ ResNet blocks with 2D convolutions, group normalization, and both upsampling and downsampling layers, further modified in~\cite{lemercier_storm_2023} to take complex valued input. 
The \emph{SGMSE+} model is trained via denoising score matching, with hyperparameters $\sigma_{min} = 0.05$, $\sigma_{max} = 0.5$, and $\gamma = 1.5$. The predictor-corrector sampler with $30$ time steps ($60$ DNN evaluations) is used during inference~\cite{richter2023sgmse}.
Under the scope of Ornstein-Uhlenbeck SDE, the \emph{SB model} uses the  Variance Exploding (VE) schedule with $\sigma_{min} = 0.7$ and $\sigma_{max} = 1.82$. For inference, SDE sampler with $50$ time steps ($50$ DNN evaluations) is used. Both SGMSE+ and SB models utilize \mbox{NCSN+} backbones with extra noise scheduling layers.
Further, exponential moving average with a weight decay of $0.999$ is used for both models~\cite{jukic2024sb}. 

In terms of model complexity, the number of trainable parameters is $7.6$M for the \mbox{MM model}, $22.1$M for the \mbox{CR} model, $25.2$M for the \mbox{SGMSE+} model, and $25.2$M for the \mbox{SB} model.

Training is carried out using a batch size of $8$ and a total of $1000$ epochs, with early stopping if the validation loss does not decrease for $20$ consecutive epochs. Furthermore, we use the Adam optimizer and a learning rate of $10^{-4}$. Training the CR, SGMSE+ and SB using the CROWD dataset was conducted on an NVIDIA H100 GPU, while all other models were trained on an RTX 3090 GPU.

\subsection{Evaluation}
The performance is evaluated through four objective measures, i.e., the extended short-time objective intelligibility (ESTOI))~\cite{estoi}, the wideband perceptual evaluation of speech quality (PESQ)~\cite{pesq_2001}, the frequency-weighted segmental SNR (fwSSNR)~\cite{fwSSNR}, and the SI-SDR~\cite{roux_sdr_2019}. For all these measures, the clean speech signal is used as the reference signal, and higher values indicate better performance. To facilitate comparison among different datasets, we report the difference of the metric values between the enhanced signals and the noisy mixtures. 

\section{Results and Discussion}

\subsection{Performance of SE models on the CROWD dataset}
\label{sec:spanish}
Since SE models are traditionally benchmarked on English data, we first evaluate performance on the neurotypical Spanish CROWD dataset to establish a baseline. This step is important because our subsequent analyses focus on Spanish, given that the available pathological speech dataset (PC-GITA) is in Spanish. Table~\ref{tbl: lang} presents the performance of the considered models when trained and tested on the neurotypical CROWD dataset.
As expected from the SOTA literature, the CR and SB models generally achieve the best performance, followed by the SGMSE+ model, while the MM model typically performs the worst. 
More specifically, the CR model yields the highest $\Delta$E-STOI, $\Delta$PESQ, and $\Delta$SI-SDR, whereas the SB model yields the highest $\Delta$fwSSNR. 
The CR and SB models will therefore be used in our subsequent analysis as representatives of SOTA predictive and generative SE models.

\begin{table}
    \centering
    \caption{ Performance of SE models trained and tested on the neurotypical Spanish CROWD dataset. Values in bold indicate the highest performance improvement obtained for each measure.}
    \label{tbl: lang}
    \addtolength{\tabcolsep}{-0.5em}
    \begin{tabularx}{.5\textwidth}{X|rrrr}
    \toprule
    Model & $\Delta$E-STOI & $\Delta$PESQ & $\Delta$fwSSNR & $\Delta$SI-SDR  \\
    \midrule
    MM & $ 0.12 \pm 0.00 $ & $1.19 \pm 0.01 $ & $ 2.55 \pm 0.04 $ & $ 9.35 \pm 0.08 $  \\
    CR & $ \bf 0.16 \pm 0.00 $ & $ \bf 1.40 \pm 0.01 $ & $ 4.13 \pm 0.04 $ & $ \bf 11.60 \pm 0.09 $   \\
    SGMSE+ & $ 0.11 \pm 0.00 $ & $ 0.75 \pm  0.01$ & $ 3.71 \pm 0.04 $ & $ 6.33 \pm 0.06 $  \\
    SB  & $ 0.15 \pm 0.00 $ & $ 1.36 \pm 0.01 $ & $ \bf 5.19 \pm 0.04 $ & $ 8.29 \pm 0.09 $  \\
    \bottomrule
    \end{tabularx}
\end{table}


\subsection{Performance of SE models on pathological speech}
\label{sec: patho}
In this section, we evaluate the CR and SB models trained on the neurotypical CROWD dataset from Section~\ref{sec:spanish} on both neurotypical and pathological speech from the PC-GITA dataset. Results are shown in Table~\ref{tbl: patho}.  
Comparing the results in Tables~\ref{tbl: lang} and~\ref{tbl: patho}, it can be observed that the performance on neurotypical PC-GITA speakers is lower than on neurotypical CROWD speakers, showing that SOTA SE models still face challenges with cross-database generalization even within the same speaker group.  
More importantly, Table~\ref{tbl: patho} shows that for both considered SE models across all evaluation metrics, performance drops even further for pathological speakers compared to neurotypical speakers. 
This gap is expected due to the different statistical distributions of neurotypical and pathological speech~\cite{PD_ITG, kodrasi_spectro-temporal_2020} and highlights a critical limitation of current SOTA models that are trained exclusively on neurotypical speech.

\begin{table*}[t!]
    \centering
    \caption{Performance of the CR and SB models trained on the neurotypical CROWD dataset and tested on neurotypical and pathological speakers from the PC-GITA dataset. Values in bold indicate the speaker group for which the highest performance improvement is obtained for each model and each measure.}
    \label{tbl: patho}
    \addtolength{\tabcolsep}{-0.4em}
    \begin{tabularx}{\textwidth}{X|cccc|cccc}
    \toprule
      & \multicolumn{4}{c|}{CR model} & \multicolumn{4}{c}{SB model} \\
    Speaker  & $\Delta$E-STOI & $\Delta$PESQ & $\Delta$fwSSNR & $\Delta$SI-SDR 
     & $\Delta$E-STOI & $\Delta$PESQ & $\Delta$fwSSNR & $\Delta$SI-SDR \\
    \midrule
     Neurotypical & $ \bf 0.09 \pm 0.00 $ &  $ \bf 0.89 \pm 0.02 $ & $  \bf 3.57 \pm 0.08 $ & $ \bf 4.22 \pm 0.19 $ & $ \bf 0.06 \pm 0.00$ & $ \bf 0.52 \pm 0.02$ & $ \bf 3.10 \pm 0.09$ & $ \bf 1.40 \pm 0.18$ \\
     Pathological& $0.05 \pm 0.00$ & $0.63 \pm 0.02$ & $2.78 \pm 0.09$ & $2.81 \pm 0.20$ & $0.01 \pm 0.00$ & $0.31 \pm 0.02$ & $2.24 \pm 0.10$ & $0.36 \pm 0.19$ \\
    \bottomrule
    \end{tabularx}
\end{table*}

\begin{table*}[t!]
    \centering
    \caption{Performance of the CR and SB models trained using various strategies outlined in Section~\ref{sec: patho_se} and tested on neurotypical and pathological speakers from the PC-GITA dataset (using a stratified cross-validation framework).}
    \label{tbl: scratch}
    \addtolength{\tabcolsep}{-0.17em}
    \begin{tabularx}{\textwidth}{X|cccc|cccc}
    \toprule
     & \multicolumn{4}{c|}{CR model} & \multicolumn{4}{c}{SB model} \\
      Speaker & $\Delta$E-STOI & $\Delta$PESQ & $\Delta$fwSSNR & $\Delta$SI-SDR 
     & $\Delta$E-STOI & $\Delta$PESQ & $\Delta$fwSSNR & $\Delta$SI-SDR \\
     \midrule
    \midrule
    & \multicolumn{8}{c}{Models trained from scratch on PC-GITA} \\
    \midrule
     Neurotypical &  $  0.15 \pm 0.00 $ & $  1.21 \pm 0.02 $ & $ 5.38 \pm 0.07 $ & $ 8.19 \pm 0.12 $ & $ 0.17 \pm 0.00 $ & $ 1.39 \pm 0.02 $ & $ 6.13 \pm 0.08 $ & $ 8.00 \pm 0.14 $ \\
     Pathological &  $ 0.13 \pm 0.00 $ & $ 1.11 \pm 0.02 $ & $ 4.94 \pm 0.08 $ & $ 7.75 \pm 0.14 $ & $ 0.15 \pm 0.00 $ & $ 1.22 \pm 0.02 $ & $ 5.47 \pm 0.09 $ & $ 7.49 \pm 0.14 $ \\
    \midrule
    & \multicolumn{8}{c}{Models trained on CROWD and fine-tuned on PC-GITA} \\
    \midrule
         Neurotypical &  $ 0.18 \pm 0.00 $ & $ 1.40 \pm 0.02 $ & $ 6.03 \pm 0.07 $ & $ 8.99 \pm 0.12 $ & $ 0.20 \pm 0.00 $ & $ 1.53 \pm 0.02 $ & $ 6.73 \pm 0.07 $ & $ 8.48 \pm 0.13 $ \\
     Pathological &  $ 0.15 \pm 0.00 $ & $ 1.25 \pm 0.02 $ & $ 5.37 \pm 0.08 $ & $ 8.29 \pm 0.13 $ & $ 0.17 \pm 0.00 $ & $ 1.31 \pm 0.02 $ & $ 5.94 \pm 0.09 $ & $ 7.66 \pm 0.14 $ \\
     \midrule
    & \multicolumn{8}{c}{Models trained on CROWD and fine-tuned with speaker-specific PC-GITA data} \\
    \midrule
     Neurotypical &  $ 0.13 \pm 0.00 $ & $ 1.11 \pm 0.02 $ & $ 4.71 \pm 0.10 $ & $ 6.63 \pm 0.21 $ & $ 0.14 \pm 0.00 $ & $ 0.70 \pm 0.02 $ & $ 5.17 \pm 0.08 $ & $ 6.20 \pm 0.13 $ \\
     Pathological&  $ 0.10 \pm 0.00 $ & $ 0.88 \pm 0.02 $ & $ 3.86 \pm 0.10 $ & $ 6.02 \pm 0.18 $ & $ 0.11 \pm 0.00 $ & $ 0.55 \pm 0.02 $ & $ 4.32 \pm 0.08 $ & $ 5.31 \pm 0.13 $ \\
    \bottomrule
    \end{tabularx}
\end{table*}

\subsection{Pathological SE models}

In this section, we explore the following strategies to improve the SE performance of the CR and SB models for pathological speakers:

{\emph{1. Training models from scratch.}} \enspace 
First, we investigate whether current SOTA SE enhancement models can be trained from scratch on small clean speech databases such as PC-GITA. This contrasts with the conventional practice of training SE models on much larger corpora such as CROWD, consisting of only neurotypical speech. 

{2. \emph{Fine-tuning pre-trained models.}} \enspace We further explore the potential of fine-tuning SE models that have been pre-trained on the larger CROWD dataset using the PC-GITA dataset.

{\emph{3. Speaker-specific models.}} \enspace We finally assess the performance of speaker-specific SE models, where the models pre-trained on the CROWD dataset are fine-tuned using data from each individual speaker from the PC-GITA dataset. To this end, we use a subset of a speaker's recordings as fine-tuning/validation data, while the remaining recordings are used as testing data. To ensure that the conclusions we draw from these models are not influenced by the specific subset of data used for fine-tuning/validation/testing, we report the average performance of two speaker-specific models, one using the monologue recording for fine-tuning/validation and the sentences and read text recordings for testing, and another vice versa. It should be noted that these subsets of data have a similar duration, i.e., the average duration of the monologue recording across all speakers is $47.1$ s, whereas the average duration of the sentences and read text recording across all speakers is $55.3$ s. 

The performances obtained using these different strategies are shown in Table~\ref{tbl: scratch} and can be summarized as follows:

{\emph{Training from scratch:}} \enspace Although the CR and SB models are relatively large, they can both be successfully trained on the relatively small PC-GITA dataset. This is confirmed by the fact that performance on neurotypical speakers is comparable to that of the same models trained on the much larger CROWD corpus (cf. Table~\ref{tbl: lang}).
Most importantly, it can be observed that including pathological data in training considerably improves performance for pathological speakers. However, a consistent gap remains between neurotypical and pathological speakers for both models. Between the two models, SB generally outperforms CR across most metrics, except for SI-SDR improvement where CR shows a slight advantage.

{\emph{Fine-tuning pre-trained models:}} \enspace Fine-tuning models trained on CROWD with pathological data yields the best overall performance, outperforming training from scratch across all metrics for both considered models. This suggests that large neurotypical corpora are valuable for learning general clean speech characteristics, while fine-tuning with relatively little pathological data enables models to capture disorder-specific traits. As with training from scratch, SB typically surpasses CR except in SI-SDR improvement. The gap between neurotypical and pathological speakers, however, persists.

{\emph{Speaker-specific fine-tuning:}} \enspace Speaker-specific models achieve the lowest performance among all strategies, across both speaker groups and all metrics. Additional experiments (not shown here due to space constraints) suggest this is due to limited amount of data available per speaker.
With only $\approx 50$ s of data per speaker, the models cannot learn robust representations, leading to poorer generalization compared to strategies that leverage more data from multiple speakers.
In summary, the most effective strategy for improving SE performance on pathological speech is to pretrain SOTA models on large neurotypical corpora and fine-tune them on smaller pathological datasets. Across these strategies, and among the considered exemplary models, SB proves to be the most effective model overall. Importantly, a performance gap between neurotypical and pathological speakers remains. We suspect this is due to the high variability in pathological speech characteristics such that training on data from other pathological speakers may not sufficiently capture the acoustic traits of a given test speaker. Future work should explore pathology-aware fine-tuning strategies that explicitly integrate domain knowledge about speech disorders and variability across individuals.
Additionally, listening tests should be conducted to confirm the conclusions derived from objective metrics.

\section{Conclusion}
This paper systematically analyzed the performance of SOTA predictive and generative SE models for pathological speakers.
While these models perform well for neurotypical speakers, their effectiveness on pathological speech remains limited.
To address this, we explored three strategies, i.e., training from scratch on pathological datasets, fine-tuning neurotypical-pretrained models with pathological data, and adapting neurotypical-pretrained models with speaker-specific data. 
Results showed that fine-tuning neurotypical-pretrained models with pathological data consistently yields the best performance for pathological speakers across both the predictive CR and generative SB models, with the SB model generally providing the strongest performance.
Nevertheless, a clear performance gap between neurotypical and pathological speakers persists, likely due to the high variability across pathological speech. These findings highlight the importance of developing pathology-aware fine-tuning strategies that explicitly incorporate domain knowledge about speech disorders, enabling SE systems to be more inclusive and effective.

\footnotesize
\bibliographystyle{IEEEtran}
\bibliography{refs}

\end{document}